# EOSC-LIFE: BUILDING A DIGITAL SPACE FOR THE LIFE SCIENCES

**EOSC-LIFE WP4 TOOLBOX:**
Toolbox for sharing of sensitive data - a concept description

WP4 – **Policies, specifications and tools for secure management of sensitive data for research purposes**
Lead Beneficiary: **ECRIN, BBMRI**
Date of publication: **30 January 2021**

Grant agreement no. 824087
Horizon 2020
H2020-INFRAEOSC-2018-2
Type of action: RIA

**Authors: Jan-Willem Boiten (EATRIS/Lygature), Christian Ohmann (ECRIN), Ayodeji Adeniran (BBMRI), Steve Canham (ECRIN), Monica Cano Abadia (BBMRI), Gauthier Chassang (INSERM), Maria Luisa Chiusano (EMBRC), Romain David (ERINHA), Maddalena Fratelli (EATRIS/IRFMN), Phil Gribbon (EU-OPENSCREEN/ Fraunhofer IME), Petr Holub (BBMRI), Rebecca Ludwig (EATRIS), Michaela Th. Mayrhofer (BBMRI), Mihaela Matei (ECRIN), Arshiya Merchant (ELIXIR), Maria Panagiotopoulou (ECRIN), Luca Pireddu (CRS4/BBMRI), Alex Sanchez Pla (EATRIS/VHIR), Irene Schlünder (BBMRI), George Tsamis (EMBRC), Harald Wagener (Charité)**

# Table of Contents

This project has received funding from the *European Union's Horizon 2020 research and innovation programme* under grant agreement No 824087.

# Toolbox for sharing of sensitive data - a concept

Authors: Jan-Willem Boiten (EATRIS/Lygature), Christian Ohmann (ECRIN), Ayodeji Adeniran (BBMRI), Steve Canham (ECRIN), Monica Cano Abadia (BBMRI), Gauthier Chassang (INSERM), Maria Luisa Chiusano (EMBRC), Romain David (ERINHA), Maddalena Fratelli (EATRIS/IRFMN), Phil Gribbon (EU-OPENSCREEN/ Fraunhofer IME), Petr Holub (BBMRI), Rebecca Ludwig (EATRIS), Michaela Th. Mayrhofer (BBMRI), Mihaela Matei (ECRIN), Arshiya Merchant (ELIXIR), Maria Panagiotopoulou (ECRIN), Luca Pireddu (CRS4/BBMRI), Alex Sanchez Pla (EATRIS/VHIR), Irene Schlünder (BBMRI), George Tsamis (EMBRC), Harald Wagener (Charité)

Status: final
Date: 30 January 2021

ABSTRACT

The Horizon 2020 project EOSC-Life brings together the 13 Life Science 'ESFRI' research infrastructures to create an open, digital and collaborative space for biological and medical research. Sharing sensitive data is a specific challenge within EOSC-Life. For that reason, a toolbox is being developed, providing information to researchers who wish to share and/or use sensitive data in a cloud environment in general, and the European Open Science Cloud in particular. The sensitivity of the data may arise from its personal nature but can also be caused by intellectual property considerations, biohazard concerns, or the Nagoya protocol. The toolbox will not create new content, instead, it will allow researchers to find existing resources that are relevant for sharing sensitive data across all participating research infrastructures (F in FAIR). The toolbox will provide links to recommendations, procedures, and best practices, as well as to software (tools) to support data sharing and reuse. It will be based upon a tagging (categorisation) system, allowing consistent labelling and categorisation of resources. The current design document provides an outline for the anticipated toolbox, as well as its basic principles regarding content and sustainability.

## 1.  OBJECTIVE & OVERVIEW OF THE DATA SHARING TOOLBOX

A large segment of the data to be handled and processed by biomedical research infrastructures is sensitive data. Sharing those data is causing an additional layer of complexity on top of the more generic technical and data security issues. The sensitivity of the data may arise from its personal nature (in particular health data), but can also be caused by intellectual property considerations, biohazard concerns, or the Nagoya protocol. In all these cases particular care needs to be taken when sharing or reusing the data, in particular in a cloud context.

The EOSC-Life Toolbox aims to provide guidance to:

1) Researchers or other data providers (e.g., sponsors, institutions or private organisations) wishing to make their sensitive data available for future reuse, *i.e., enabling future sharing of data*;
2) Researchers and data providers in the case where a researcher wishes to make use of sensitive data made available by the data provider*, i.e., enabling actual sharing of data*.

The content of toolbox will be derived from two main sources: (1) existing recommendations, procedures, best practices, and links to software (tools) to support data sharing and reuse operations relevant for sensitive data management in the cloud environment, esp. EOSC; (2) guidelines and other useful resources drafted in the context of EOSC-life (e.g., WP7 guidelines on the required capabilities of cloud providers). The toolbox will not be designed *de novo*; instead, it will help scientists to navigate to previously collected high-quality content available throughout our collective infrastructure landscape. Also, the sustainability of the toolbox will be a design consideration from the start: EOSC-Life is a project with a clear end date. Any toolbox can only be useful and successful if it is maintained and updated beyond that end date.

The creation of the toolbox and filling it with content will be driven by use cases, which may stem from within EOSC-Life (e.g., WP1, WP3, WP6 (Nagoya protocol)) or from the Research Infrastructures' network / other EU-projects. Data sharing issues highlighted by those use cases will prioritize the content collection for the toolbox and will provide test cases for the usability of the toolbox.

The toolbox for data sharing will be the main output for the EOSC-LIFE deliverable 4.3 ("Guidance and policy on standards and tools to facilitate sharing and reuse of multimodal data (including imaging), cohort integration and biosamples").

## 2.  SCOPE OF THE TOOLBOX

Control of the scope of the toolbox is essential in order to be able to create a useful product within the time and budget available. A high-level overview of this scope in a few bullets:

- Only secondary use of data is considered; in case of primary use only those aspects related to future reuse of the data will be considered.

This project has received funding from the *European Union's Horizon 2020 research and innovation programme* under grant agreement No 824087.

- The main actor for the toolbox is the researcher with two viewpoints:
    - a researcher who is sharing data or wishes to share the data in the future, or
    - a researcher who is requesting data for secondary use.
- A certain level of expertise is expected from the targeted audience, i.e., "intermediate level of expertise" is expected: neither a novice in data sharing nor an ethical/legal expert. It should also be considered that users may be experienced in one aspect, while being novice in other areas. Although not really in scope for the toolbox, we do intend to provide some resources for novice users.
- The content of the toolbox should be of relevance for cloud usage of sensitive data in general, and the European Open Science Cloud in particular. The priority for filling the toolbox will be set by either:
    - representative use cases, or
    - specific topics raised by other EOSC-Life work packages, or
    - specific input from the WP4 EOSC-Life toolbox team.
- The toolbox can consist of recommendations, procedures, links to tools, and examples/best practices. Preferably the best practices are derived from the use cases.
- The toolbox will not be developed *de novo*, but assembled from existing resources, preferably by providing pointers to resources rather than copying content. However, the toolbox will provide sufficient context to those pointers to allow users to understand their scope and applicability.

It would be a pitfall to aim for the perfect product from the start. The initial focus should be the "minimum lovable product".

## 3. GENERAL OUTLINE OF THE TOOLBOX

The guidance for handling sensitive data in EOSC-Life will be provided via a toolbox where we have three main aspects to consider:

1. Use cases/user stories
2. Content: guidances, tools, and best practice description + some generic background information for novice users
3. Navigation, user interface elements and functionalities to easily access the content

The toolbox is a website, where different navigation options should guide users to the right content, not only for simple use cases basically consisting of one data sharing issue, but also for more complex use cases having to deal with multiple issues. Maintenance and sustainability of the website will be an essential design consideration: it should be easy to keep the content up-to-date and the responsibility beyond the EOSC-Life project lifetime should be clear from the start. The value of the toolbox highly depends on the reliability and actuality of the content, not only for the toolbox itself, but also for the content from other sources that are referenced. A mechanism to keep control of the actuality of that content should be considered.

A disclaimer should be included on the toolbox declining all warranties in case recommendations from the toolbox have caused any damages.

## 4. USE CASES & USER STORIES

Prototypic use cases are identified that represent the data sharing needs of our research community. The use cases should be well selected in order to be sufficiently specific to be easily translatable to concrete user needs in practice, while at the other hand the use cases should be sufficiently generic to allow for broad usage. So, it is required to have a representative set of use cases covering a broad range of data sharing problems related to EOSC-Life. The use cases can be derived from the existing and upcoming demonstrators in WP1, 3, and 6, and from specific topics raised by other EOSC-Life work packages.

Concrete data sharing dilemmas will be distilled from these use cases, maintaining the link to the use cases to make the needs tangible. Different approaches will be used to represent the use cases (e.g., textual user stories or forms to be filled in) and to link the use cases to the content.

Currently we have a number of sources of use cases which can be included:

1. The evaluation of the applications for the WP1 open call for use cases.
2. Common use cases with WP2 allowing to iron out the requirements for handling sensitive data in the toolset developed in that work package, in particular for workflows to be used for sensitive data. At this time, only one use case with sensitive data has been identified by WP2: Sharing digital pathology images (led by Petr Holub).
3. The -Beyond One Million Genomes- project (https://b1mg-project.eu/) could offer a valuable source of use cases. Four use case areas have been identified in this project: rare disease, cancer, common complex disease, and infectious disease/COVID-19. Each of these areas need to be more specific to be suited for guiding the toolbox development.
4. A set of use cases previously collected: (Paris; January 23rd, 2020) as described in a separate working document[1].

These use cases should be as concrete as possible: real research projects with real data sharing issues. When generalizing too soon we may miss the real-life problems which are often caused by apparently insignificant details.

From these use cases we can derive more generic user stories. This can be done in multiple ways: in the Paris meeting referred to above, we tried to identify the user stories as generic data sharing issues. This has been further prioritized in two sessions as summarized in the minutes[2] from the EOSC-Life WP4 meeting in Brussels (March 4th, 2020). Alternatively user stories can be summarised in simple statements from the researcher's perspective in an extensive list of examples collected by ECRIN (Steve Canham). These user stories are all formatted in a similar fashion, e.g., "*as a researcher, as part of study planning, I want to ensure that consents I obtain will support the agreed data sharing policy*".

This project has received funding from the *European Union's Horizon 2020 research and innovation programme* under grant agreement No 824087.

## 5. CONTENT OF THE TOOLBOX

Regarding the content of the toolbox there are two main workflows to be considered which may also be reflected in the user interface:

1. Planning data sharing in a research project, i.e., planning for future sharing
2. Requesting and sharing data for secondary use, i.e., actual data sharing

### 5.1. WORKFLOW 1: PLANNING FOR SHARING

This workflow is summarized in the following workflow diagram:

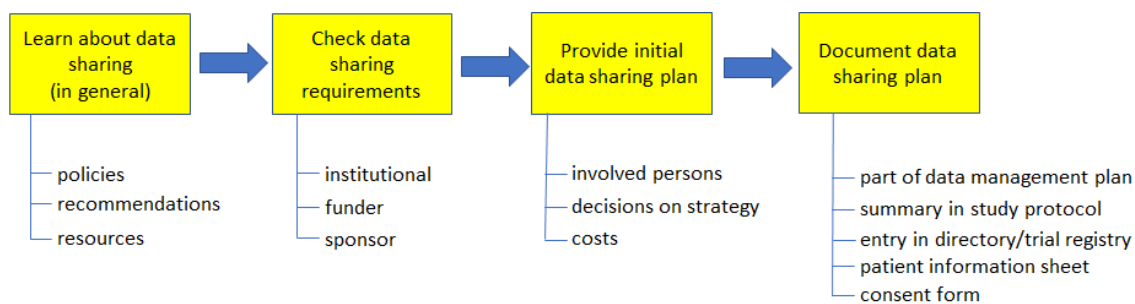

*Figure 1*: *workflow when planning for future data sharing of sensitive data.*

Possible aspects to consider when planning for future sharing:

- learn about data sharing in general (procedures, options, repositories)
- check funder/institutional requirements for data sharing
- define data sharing procedures in data management plan
- provide summary of data sharing procedure in study/research protocol
- informed consent
- incorporate data sharing summary in study registry (if relevant)
- align data sharing plan with all collaborators of the project
- learn about ethical and legal frameworks

### 5.2. WORKFLOW 2: REQUESTING AND SHARING DATA

Possible aspects to consider when actually wishing to share data:

- if data sharing is planned prospectively, data sharing plan needs to be reviewed and updated or otherwise a new data sharing plan needs to be developed
- decision on de-identification (e.g., anonymisation, pseudonymisation - dependent on explicit consent)
- application of de-identification

- carry out risk assessment: anticipate risks and plan mitigation measures (e.g., Data Protection Impact Assessment (DPIA) if requested)
- select file formats for data and meta data
- explore repository, where the data should be stored (conditions, access tape, costs)
- sign data transfer agreement
- transfer data to sustainable repository
- monitor use of transferred data
- submit request (form) to repository
- sign data use agreement with repository
- download requested data on your computer or use protected analysis environment
- describe data sharing tools / conditions / contact / licences / provenances in a data paper
- perform secondary analysis
- publish secondary analysis
- citing the data paper in the secondary publication

Figure 2 aims to visualize these aspects in a workflow:

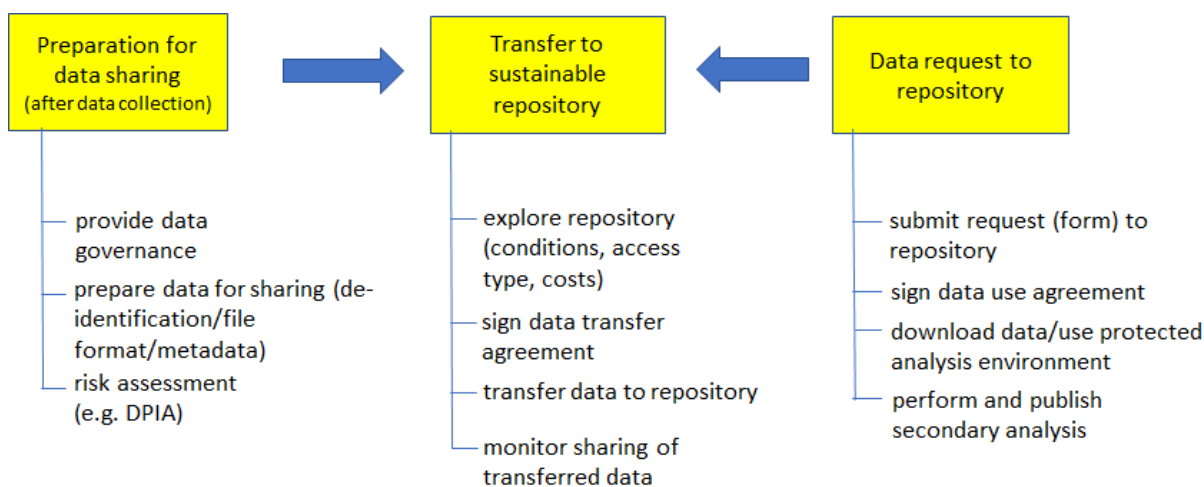

**Figure 2**: Key aspects to consider when making sensitive data available for sharing.

It is not the intention of the EOSC-Life project to develop new guidelines, best practices, tools/services. Instead, existing content should be assessed according to standardised criteria relevant for the EOSC-Life project. However, in case of certain white spots, EOSC-Life WP4 may organize expert workshops to gain access to specific content for the toolbox. An example of such a workshop was the meeting on pseudonymization/anonymization organized in Paris (January 22-23, 2020). The draft report[3] of this meeting is available.

The BBMRI-ERIC Knowledge Base (as described in a publication[4]) may provide an important starting point for the design, relevant content, and sustainability aspects of the toolbox:

This project has received funding from the *European Union's Horizon 2020 research and innovation programme* under grant agreement No 824087.

https://www.bbmri-eric.eu/elsi/knowledge-base/. AA demonstration of that Knowledge Base for this working group was organized July 27$^{th}$, 2020. A new release of the Knowledge Base was issued in autumn 2020. The next version update is scheduled for Q2 2021.

EOSC-Life WP7 is currently working on a guideline for the required capabilities of cloud providers when wishing to process sensitive data. The resulting guidelines should certainly be included in the toolbox.

## 6. NAVIGATION AND USER INTERFACE

The navigation is essential to the use of the toolbox and should be based upon the needs of a user related to a specific problem or use case. EOSC-Life WP9 is supporting the usability aspects of the toolbox implementation ("user journey"). A usability webinar (September 28$^{th}$, 2020) has been organized to inform the toolbox team regarding tips and tricks in this realm. It was based on the EBI Training on user experience: https://www.ebi.ac.uk/training-beta/online/courses/user-experience-design/what-is-user-experience-design/

One possible methodological approach could be to follow the approach outlined in the book by J.J. Garrett[5] distinguishing product as a functionality and product as information, covering different planes.

Another example that could inspire us with the further creation of the toolbox is the Data Stewardship Wizard created for the RDA COVID-19 initiative. A demo of this toolbox was organized for the larger EOSC-Life Toolbox working group on August 24$^{th}$, 2020.

We will start a small working party that is dedicated to the technology to be used for the toolbox. The selected technology should be based on open-source software to avoid vendor lock-in (whenever possible). Google Analytics is an important tool to be used to obtain an indication of the actual usage of the toolbox.

A few generic considerations for the navigation:

- In terms of usability, cross-references should be used with care. When too many clicks are involved users tend to give up.
- Management of the content should be independent of the navigation. It should be possible to update the content without breaking the navigation.

Multiple navigation approaches need to be supported; possible options are:

- Decision tree; it has the clear advantage that steps are executed in the right order
- Direct search
- Infographic
- Classic alphabetical index
- Landing page with most recent additions
- A toolbox in a nutshell page with most useful / used information

- A hide/show button for each functionality to permit users to personalize /simplify the interface

The actual UX design should ideally be tested in practice with representatives from the intended user group. Also, an on-line feedback option could serve as a source of user feedback during the production phase of the toolbox. It may also be a way to capture reports on outdated content.

Direct search may be supported by tagging the content. This basic approach to navigation could be supported by a **form/questionnaire** to be filled in, which characterises the main dimensions of data sharing. The content (e.g., guidances, best practice, tools) could be tagged with these characteristics and thus made identifiable via search. A detailed proposal[6] for such a tagging system has been created, as well as a study protocol[7] to test its practical usability. This could be a first higher layer for discovery of content, which may be extended by more specific navigation, if needed. A question to be answered is whether and how the different navigation strategies (wizard, mind-mapping, direct search, form) could work together and whether tagging/annotation of content can be used to serve different navigation strategies. In fact, a hybrid approach combining different techniques will probably be preferable.

## 7. SUSTAINABILITY & TOOLBOX GOVERNANCE

The usefulness of the toolbox is not only determined by the usability of the toolbox, but also by the quality, the reliability, and the actuality of its content. It is therefore crucial to have a (lightweight) governance process in place, e.g., by reviewing every toolbox topic by two or three independent experts before publishing it. One of the review criteria should cover the predicted sustainability of the content. Previous experience shows that updates of documents are often not officially released anymore[8]. It should be unlikely that the content will be outdated rapidly; nevertheless, some periodic review will be needed to maintain the content of the toolbox over time. Preferably, this should be automated as far as possible, but experience from BBMRI's Knowledge Base tells that this is hard to realise in practice. For that reason, BBMRI is using a manual 3-monthly update cycle for the content on the ELSI service page on the BBMRI-ERIC website.

## 8. REFERENCES

1. Boiten, J.W. Guidance for sensitive multimodal data in EOSC-Life – minutes of first scoping discussion, January 23rd, 2020, Paris.
   https://drive.google.com/file/d/1LEHK3WVuSEcbUCKvsmmIMNBR5tnlJecU/view?usp=sharing

This project has received funding from the *European Union's Horizon 2020 research and innovation programme* under grant agreement No 824087.

This project has received funding from the *European Union's Horizon 2020 research and innovation programme* under grant agreement No 824087.